\documentclass[12pt,preprint]{aastex}
\usepackage{graphics}
\begin{document}

\newcommand{\kms}{\mbox{km~s$^{-1}$}}
\newcommand{\mloss}{\mbox{$\dot{M}$}}
\newcommand{\my}{\mbox{$M_{\odot}$~yr$^{-1}$}}
\newcommand{\ls}{\mbox{$L_{\odot}$}}
\newcommand{\ms}{\mbox{$M_{\odot}$}}
\newcommand\mdot{$\dot{M}$}
\def\arcdeg{\hbox{$^\circ$}}
\def\farcs{\hbox{$.\!\!^{\prime\prime}$}}
\def\gtrsim{\mathrel{\hbox{\rlap{\hbox{\lower4pt\hbox{$\sim$}}}\hbox{$>$}}}}
\def\lesssim{\mathrel{\hbox{\rlap{\hbox{\lower4pt\hbox{$\sim$}}}\hbox{$<$}}}}

\title{AN EVLA AND CARMA STUDY OF DUSTY DISKS AND TORII WITH LARGE GRAINS IN DYING STARS}
\author{R. Sahai\altaffilmark{1}, M. J. Claussen\altaffilmark{2}, S. Schnee\altaffilmark{3}
 M. R. Morris\altaffilmark{4}, C. S{\' a}nchez Contreras\altaffilmark{5}
}

\altaffiltext{1}{Jet Propulsion Laboratory, MS 183-900, California
Institute of Technology, Pasadena, CA 91109}

\altaffiltext{2}{National Radio Astronomy Observatory, 1003 Lopezville Road,
Socorro, NM 87801}

\altaffiltext{3}{National Radio Astronomy Observatory, 520 Edgemont Road, Charlottesville, VA 22903}

\altaffiltext{4}{Division of Astronomy, Department of Physics and
Astronomy, UCLA, Los Angeles, CA 90095}

\altaffiltext{5}{Astrobiology Center (CSIC-INTA), ESAC campus, E-28691 Villanueva de la Canada, Madrid, Spain}

\email{raghvendra.sahai@jpl.nasa.gov}

\begin{abstract}
We report the results of a pilot multiwavelength survey in the radio continuum (X, Ka and Q bands, i.e., from 3.6 cm to 7 mm)
carried out with the EVLA in order to confirm the presence of very large dust grains in dusty disks and torii around the central
stars in a small sample of post-AGB objects, as inferred from millimeter and sub-millimeter observations. Supporting mm-wave
observations were also obtained with CARMA towards three of our sources. Our EVLA survey has resulted in a robust detection of
our most prominent submm emission source, the pre-planetary nebula IRAS\,22036+5306, in all three bands, and the disk-prominent
post-AGB object, RV\,Tau, in one band. The observed fluxes are consistent with optically-thin free-free emission, and since they
are insignificant compared to their submm/mm fluxes, we conclude that the latter must come from substantial masses of cool, large
(mm-sized) grains. We find that the power-law emissivity in the cm-to-submm range for the large grains in IRAS22036 is
$\nu^{\beta}$, with $\beta=1-1.3$. Furthermore, the value of $\beta$ in the 3 to 0.85\,mm range for the three disk-prominent
post-AGB sources ($\beta\le0.4$) is significantly lower than that of IRAS22036, suggesting that the grains in
post-AGB objects with circumbinary disks are likely larger than those in the dusty waists of pre-planetary nebulae.
\end{abstract}


\section{Introduction}
Ordinary stars (i.e., with main-sequence masses 1-8\ms) die extraordinary deaths, producing profound effects on their
environment before they fade into obscurity as white dwarfs. The mass-loss phenomena in these objects play a key role
in the Galaxy's chemical and dynamical evolution, and dramatically alter stellar evolution. Post-AGB (pAGB) objects 
are believed to be objects in transition between the AGB and planetary nebula (PN)
evolutionary phases and hold the key to some of the most vexing problems in our understanding of these very late stages
of evolution for ordinary stars. 

We have made considerable progress in understanding the mass-loss processes which govern the transformation of the
spherical mass-loss envelopes of AGB stars into strikingly aspherical PNs (Balick \& Frank 2002, Sahai et al. 2011),
specifically with the identification of fast, collimated outflows as the most likely primary agent for the change in
morphology (Sahai \& Trauger 1998). But the presence of dusty disks (or highly-flattened equatorial structures) around
pAGB stars remains an enigma: an intriguing feature of these pAGB objects is the presence of strong submillimeter
excesses, which suggest the presence of very large, cold grains. Here we report observations that confirm the presence
of large grains in two key objects, paving the way for detailed dust-radiative transfer models to accurately constrain
the grain sizes and theories for the formation and longevity of disks in which grains can grow to such large sizes.  

There are two major classes of pAGB objects having equatorially-flattened structures. The first class has strong
evidence for medium-sized ($\sim$50\,AU) disks and very seldom shows prominent nebulosity (hereafter dpAGB, i.e.,
disk-prominent pAGB objects). The near-to-far infrared spectral-energy-distributions (SEDs) of these objects have been
modelled as arising from disks (de Ruyter et al. 2005, 2006); the disks are believed to be gravitationally bound. A
large fraction of the central stars of dpAGB objects consists of short-period radial-velocity binaries, and many of
them show stellar photospheric depletion patterns resulting from a poorly understood process in which the circumstellar
dust is trapped in a disk, and the dust-depleted gas is accreted back onto the star (Waters et al. 1992). Recent near-
and mid-infrared interferometric imaging with the VLTI of a subsample of these objects has provided direct evidence for
the disk structures and their sizes (e.g., van Winckel et al. 2008), confirming their circumbinary nature. The ISO-SWS
($\sim$2.5-40\micron) and Spitzer spectra of many of these show amorphous and crystalline silicate features, indicating
significant dust processing, which is only possible in a stable circumstellar disk. de Ruyter et al. (2006) modelled
the 0.85\,mm fluxes in dpAGB objects as arising from very large (mm-sized) grains; detailed 2D disk-modelling by Gielen
et al. (2007) has lent further support to this conclusion. 

The second class of pAGB objects consists of pre-planetary nebulae (PPNs), observationally recognized by prominent
aspherical nebulosities surrounding a central pAGB star. Most bipolar or multipolar PPNs harbor overdense, dusty
equatorial waists that are much larger than the dpAGB circumbinary disks. In an extensive HST imaging survey of PPNs,
Sahai et al. (2007a) recognize these waists as an important morphological feature in PPNs: for many of these, the
waists display sharp outer (radial) edges, with radii typically $\gtrsim 1000$\,AU.
For a few PPNs for which submm data are available, the latter indicate the presence of large grains as in the dpAGB
objects. For example, in the bipolar PPN IRAS\,22036+5306, Sahai et al. (2006), find an unresolved ($<$0\farcs85) submm continuum
source associated with the central, dusty torus seen in HST images (Sahai et al. 2003). Modelling of the full SED, including the
submm flux,
shows that the latter is produced by a very substantial mass ($\gtrsim$0.02\ms) of large (mm-sized), cool
(T$\lesssim$50\,K) grains. Similar results have been inferred for other PPNs, e.g., IRAS\,18276-1431 and
IRAS\,19475+3119, from OVRO 2.6\,mm data (S{\' a}nchez Contreras et al. 2007, Sahai et al. 2007b), and CRL\,2688 from
VLA 1.3--3.6\,cm data (Jura et al. 2000). 

The connection between dusty waists in PPNe, the disks in dpAGB objects, and inner accretion disks is unknown. Yet,
these
structures appear to be key ingredients in our understanding of the late evolution of most stars -- the inner disks are
the only ones that are likely to be able to contribute to jet production, while the dusty waists are the only ones that
contribute to the observed morphologies in PPNs. It is believed that accretion disks may form from either (a) the
shredding of a low mass companion around an AGB primary core (Reyes-Ruiz \& Lopez, 1999), or (b) Bondi-Hoyle accretion
of the primary wind around the companion (Morris 1987, Soker and Livio 2004). Smoothed-particle hydrodynamics (SPH) codes have
demonstrated how an
unbound outflow from a mass-losing primary can be concentrated toward the system's equatorial plane into a small,
$\sim$(1-few)\,AU, bound disk around a binary companion (Mastrodemos and Morris 1999). The Red Rectangle (HD\,44179),
which is the closest known PPN with a binary central star, provides a connecting link between these two classes of pAGB
objects -- direct evidence for a large (outer radius $\sim300-550$ AU), bound disk in this object comes from
interferometric CO(2-1) mapping (Bujarrabal et al. 2003), and HST imaging reveals a bipolar PPN (Cohen et al. 2004).

The disk or torus mass can provide an important constraint on the formation process. For example, common envelope
evolution (Moe \& de Marco 2006) would lead to expulsion of most of the stellar envelope. Hence, a large value of the
disk or torus mass ($\sim$0.01\ms~in dust), as, e.g., in IRAS22036, would support a common envelope origin;
wind-accretion would lead to much lower values for this mass (as, e.g., found in HD\,44179). However, since the
frequent detection of strong H$\alpha$ emission (e.g., Maas et al. 2005, S{\' a}nchez Contreras et al. 2008) in pAGB objects shows
that some 
ionized gas is typically present, the contribution of free-free emission from ionized gas to the
submm/mm fluxes must be determined before reliable masses can be observed. Furthermore, the dust power-law emissivity
index in the mm-to-cm wavelength range is poorly known. In order to address both these issues, we carried out a pilot
multifrequency (43.3, 33.6, \& 8.5\,GHz) survey of 10 pAGB objects with the National Radio Astronomy Observatory's\footnote{NRAO
is a
facility of the National Science
Foundation operated under cooperative agreement by Associated Universities,
Inc.} EVLA facility and report our results in this Letter.

\section{Observations} 
We selected 10 post-AGB objects (Table\,\ref{fluxes}) for EVLA observations on the basis of their published mm or submm continuum
fluxes (e.g., Gledhill et al. 2001, G\"urtler et al. 1996) and a new interferometric survey with OVRO (S{\' a}nchez Contreras \&
Sahai 2011) and CARMA (reported in this paper).

The EVLA observations were performed with the
array in the {\bf D} configuration in a series of short scheduling blocks under 
project code AS1021, from March 26, 2010 through June 21, 2010. The EVLA WIDAR correlator was
configured to give two subbands each with a bandwidth of 128\,MHz and 
dual polarization.  Three frequency bands were observed for most sources
(though not all, because of time constraints) with center frequencies of:  8396 \& 8524\,MHz (X band),
33496 \& 33624\,MHz (Ka band), and 43216 \& 43344\,MHz (Q band).  Observations
in the higher frequency bands typically had more time on source in order
to have similar signal-to-noise ratios in all three bands.  For low declination
sources, however, the increase in the rms noise levels for the two high
frequency bands are still apparent.   The synthesized beam sizes depend
upon the receiver band as well as the declination: they range from 
8\farcs2$\times$2\farcs5\,to 33\farcs3$\times$20\farcs9\,for X band; 
2\farcs2$\times$1\farcs8\,to 9\farcs5$\times$1\farcs8\,for Ka band; and 1\farcs5$\times$1\farcs4\,to
6\farcs2$\times$1\farcs7\,at Q band, depending upon the declination.

Absolute flux calibration was determined by observing either 3C286 or 3C48, which are EVLA 
flux-density standards.  The flux density scale
is likely accurate to about 5\% for X-band, high declination sources,
but degrades to 20-30\% for Q-band, and especially the low-declination
sources.  The gain (phase and amplitude) calibration was carried out
in the normal manner after initial bandpass and delay calibration were
made using the flux-density calibrators.  All calibration and imaging
steps were performed using natural weighting with the AIPS software package; fluxes were derived from Gaussian fits to the
observed intensity maps, using the JMFIT task.

Observations with CARMA (Combined Array for Research in Millimeter-wave Astronomy: a 15-element interferometer
with nine 6.1 meter dishes and six 10.4 meter dishes), covered the three dpAGB objects AC\,Her, U\,Mon \&
RV\,Tau with $\sim$(2--3)\arcsec resolution; AC\,Her \& U\,Mon were observed at 3 and 1.3\,mm whereas RV\,Tau was
observed at 3\,mm only.  
Each source was observed in single pointings, with 15-20 min spent on source and 3-5 min spent on the calibrator
in each observing cycle. A passband calibrator was observed in each track, and pointing was done every 2-3 hr for
the 1-mm observations and at the beginning of each 3-mm track. Absolute flux calibration was done using the gain and passband
calibrators in the continuously-updated CARMA flux catalog, supplemented by
observations of primary flux-density calibrators in those tracks in which such a source was available. Based on the
repeatability of the quasar fluxes, we believe that our source fluxes are accurate to within $\pm$ 20\%. Calibration
and imaging was done using the MIRIAD data reduction package, using natural weighting and the standard clean algorithm; 
fluxes were derived from Gaussian fits to the observed intensity maps, using the IMFIT task. Some of the 1.3\,mm observations were
taken in shared-risk time in which only the nine 6.1-m antennas
were in the array.

\section{Results} 
We detected 2 objects with the EVLA: IRAS22036 was detected in all 3 bands, whereas RV\,Tau was detected in the Ka band only
(Table\,\ref{fluxes}). For IRAS22036, the fluxes vary only weakly over the wavelength range covered by the X, Ka and Q bands,
i.e., from 3.5 to 0.67\,cm (Table\,\ref{fluxes}), implying that these arise predominantly from optically-thin free-free emission. 
If, for the emitting region, we assume a typical ionized gas temperature of 10$^4$\,K, and a cylindrical geometry with a length
along the line-of-sight equal to its projected lateral extent, then the observed free-free flux, $S_{ff}$, is a function of
$\theta_{ff}$ (source angular size), $n_e$ (the electron density, assumed to be uniform), and $\nu$. Thus, from the
observed 33.6\,GHz flux we derive a free-free optical depth of $\tau_{ff}$(33.6\,GHz)$>0.0002$ and $n_e>1.1\times10^4$\,cm$^{-3}$,
assuming that $\theta_{ff}<$0\farcs85 (i.e., the upper limit for the size of the submillimeter continuum source). By requiring
$\tau_{ff}$(8.5\,GHz)\,$<1$, the observed 8.5\,GHz flux implies $n_e<6.4\times10^5$\,cm$^{-3}$ and $\theta_{ff}>$0\farcs067
(130\,AU); the $\nu^{-2.1}$ dependence of $\tau_{ff}$ implies $\tau_{ff}$(33.6\,GHz)$<0.055$. The
free-free emission is unlikely to arise from an ionized outflow as it would produce a spectral variation
S$_{\nu}\propto\nu^{0.6}$ (Wright \& Barlow 1975).

For RV\,Tau (not observed in X-band), our Ka-band detection and Q-band upper limit ($107\,\mu$Jy) are inconsistent with
optically-thick free-free emission and thermal dust emission, as these produce a spectral index, 
$\alpha=d$\,lnS/$d$\,ln$\nu\ge$2, implying a Q-band free-free flux $S_{ff}(0.67\,cm)\gtrsim490\,\mu$Jy, if scaled from the Ka-band
flux. For
optically-thin free-free emission, with $\alpha=-0.1$, we expect $S_{ff}(0.67\,cm)=262\pm50\,\mu$Jy, consistent with our
3$\sigma$ upper limit of $321\,\mu$Jy. 

Mm-wave continuum emission was detected in the 3 sources observed with CARMA, AC\,Her, U\,Mon \& RV\,Tau (Table\
\ref{fluxes}), and was unresolved (beam sizes were typically $\lesssim\,2${''}), consistent with these arising in a
compact structure, likely a disk, rather than an extended outflow.

If we scale the radio-to-mm/submm flux ratios of IRAS22036 for PPNe and RV\,Tau for dpAGB objects, 
we find that the expected radio fluxes for several pAGB objects (Ka-band flux=2.45, 1.68, 1.54  and 1.04\,mJy, respectively, for
IRAS16342, 18135, 18276 and U\,Mon; X-band flux=7.21\,mJy for IRAS19548, 20000) are far above their observed upper limits (Table
1). This result supports the idea that the physical mechanism predominantly responsible for the radio emission in pAGB objects is
unrelated to that responsible for the mm/submm emission.

\begin{figure}[htbp]
\resizebox{0.77\textwidth}{!}{\includegraphics{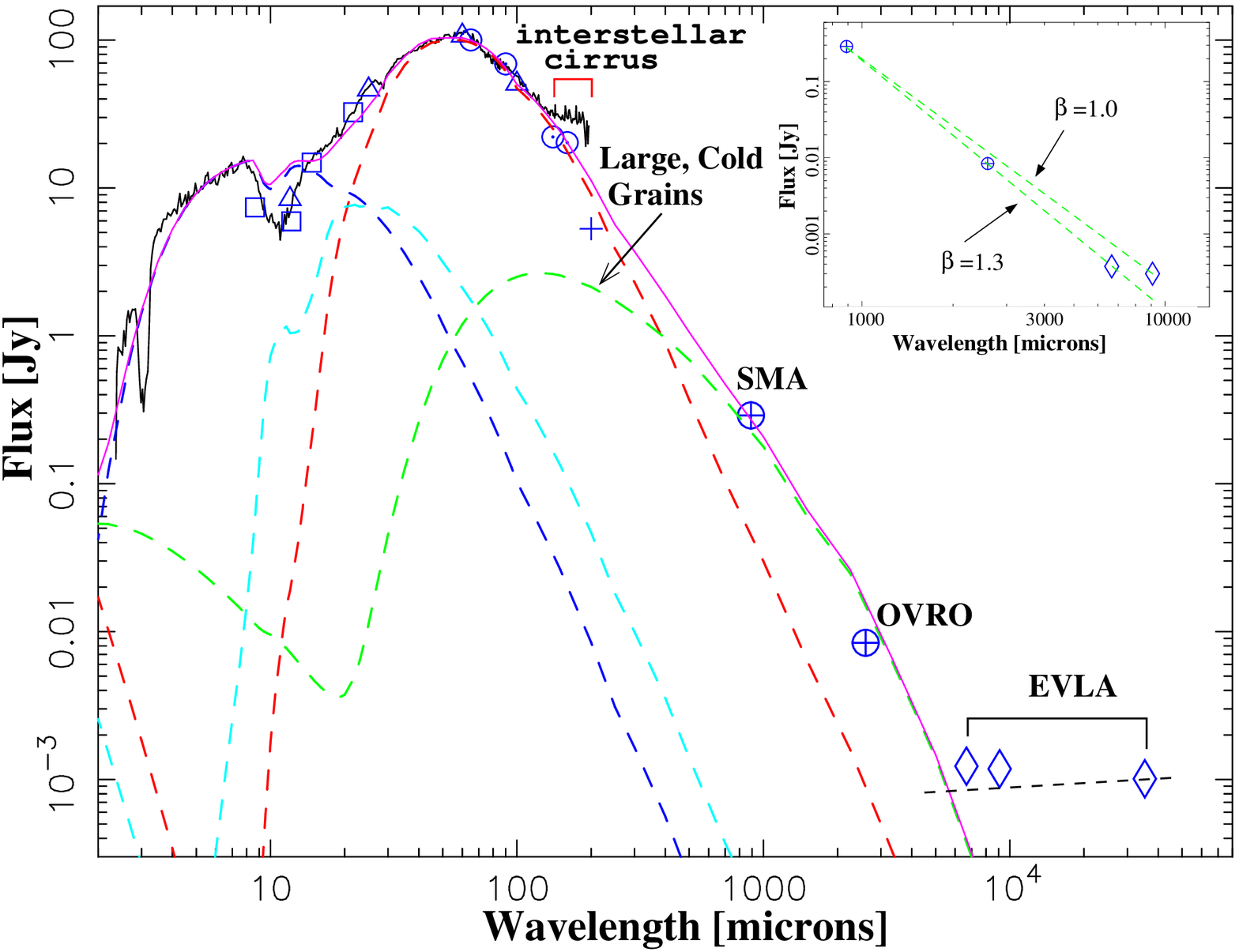}}
\caption{Observations of IRAS\,22036 [{\it black curve}: ISO spectra; {\it blue symbols}: photometric data -- {\it
squares}: MSX; {\it triangles}: IRAS; {\it circles
w/center}: AKARI; {\it cross}: ISO/PHOT; {\it circles
w/crosses}: SMA 0.88\,mm and OVRO 2.6\,mm; {\it diamonds}: EVLA 0.67, 0.9, \& 3.5 cm] and
model spectra ({\it magenta curve}). Individual components ({\it
dashed curves}) of the model are also shown: cool ({\it red}) \& warm ({\it
cyan}) shells, and hot inner disk ({\it blue}). A dashed ({\it black}) straight line shows a $\nu^{-0.1}$ power-law,
representing an optically-thin free-free fit to the 3.5\,cm EVLA flux. Inset shows the long-wavelength (submm-to-cm)
emission with the 
free-free contribution subtracted, and optically-thin dust model fits (using the Rayleigh-Jeans approximation) with two
values of the power-law dust-emissivity index, $\beta$.
}
\label{i22036}
\end{figure}

Since the observed EVLA fluxes in IRAS22036 and RV\,Tau are insignificant compared to their submm/mm fluxes, the latter
cannot be due to free-free emission and must be due to thermal emission from grains, thus providing strong support for
the presence of substantial masses of cool large grains in these objects. 
Given the very low estimated free-free optical depth in IRAS22036, we extrapolate the observed X-band flux using a
spectral index of $-0.1$, to estimate the free-free flux in the Ka and Q bands of 880 and $856\,\mu$Jy -- the excess
fluxes of 300 and $374\,\mu$Jy above the free-free extrapolated emission in these bands must then represent the long-wavelength
tail of the large grain emission
seen in the submm-mm wavelength range. In Fig.\,\ref{i22036} we show the full SED of IRAS22036, together with the dust
radiative transfer model of Sahai et al. (2006), which includes a component of cold (50\,K), large ($a=1$\,mm) grains.
The model, which uses the DUSTY code (Ivezic et al. 1999) and optical constants for silicate dust provided by Ossenkopf
et al. 
(1992), was constrained by the long-wavelength emission up to 2.6\,mm -- we find that it reproduces the excess Q-band
flux of $374\,\mu$Jy, but the predicted Ka-band flux is a factor 3.7 times lower than our value of 
the excess, $300\,\mu$Jy. This is likely due to the inadequacy of the current dust optical constants for computing dust emissivity
at cm wavelengths. We have therefore made new optically-thin dust model fits to the SED from 0.88\,mm to 0.67\,cm, using
the Rayleigh-Jeans (R-J) approximation and a power-law dust emissivity, $\kappa(\nu)\propto\nu^{\beta}$,and find that values of
$\beta$ in the range $1-1.3$ provide an adequate fit (Fig.\,\ref{i22036} inset).

We find that $\beta=0$ for RV\,Tau in the 3 to 0.85\,mm wavelength range (computed in the R-J approximation), significantly lower
than the corresponding value ($\beta=1.3$) for IRAS\,22036 (over the same wavelength range). If we assume that, like RV\,Tau, the
other two dpAGB
objects, U\,Mon and AC\,Her, also do not have significant free-free contributions at mm/submm wavelengths, then their 3\,mm
to 0.85\,mm flux ratios yield similarly low values of $\beta$ (0 and 0.4, respectively). This difference suggests that
the grains in the disks of dpAGB objects are likely larger than those in the dusty waist of IRAS\,22036, and may be
several mm in size: e.g., Draine (2006)
finds that, for the standard power-law distribution of grain sizes that characterizes interstellar dust and that may apply
to particles growing by agglomeration in protoplanetary disks, $\beta\lesssim 1$ (at $\lambda\sim1$\,mm) results when grains grow
to sizes $a\ge3$\,mm. 

We estimate the mass of the large grains ($M_d$)
for the 4 sources in Table\,\ref{fluxes} for which the observed submm flux is known to come from a compact central source,
i.e, the 3 dpAGB sources and IRAS22036, by using the Rayleigh-Jeans approximation for grains emitting at wavelength $\lambda$,
$M_d=(S_{\nu}\,\lambda^2\,D^2)/(2\,k\,T_d\,\chi_{\nu}$), where D is the source distance, and $\chi_{\nu}$ is the dust opacity
(per unit dust mass). We have not estimated dust masses for the six sources in Table\,1 for which we are
unable to associate the observed mm/submm fluxes with a central dusty torus (all of which are PPNe), either because the mm/submm
observations lack
adequate angular resolution (IRAS16342, 17150, 18135, 18276, 20000), or because there is lack of optical imaging to inform us of
the object's morphological structure (IRAS19548).  

The value of $\chi_{\nu}$ at submm wavelengths is uncertain by at least a factor of a few, and likely more. In the
extensive study of the composition and radiative properties of
grains by Pollack et al. (1994: hereafter Petal94), only large grains with $a\ge$3\,mm have values of $\beta\le1.3$ in
the 650$\micron$ to 2.7\,mm wavelength range. If we use the largest values of $\chi_{\nu}$ estimated for such grains
from this study, then setting $\chi_{\nu}(1\,mm)$ equal to $g_{d}\,\kappa_t$, where $g_{d}=71.4$ is the gas-to-dust
ratio estimated from Table\,2 of Petal94, and $\kappa_t=7.4\times10^{-3}$ and $\beta=1.34$ is taken from Table\,4 of Petal94, we
extrapolate
$\chi_{\nu}(1\,mm)$ to obtain $\chi_{\nu}(0.85\,mm)=0.65$ cm$^{2}$g$^{-1}$. 
There are no grains in the Petal94 study
with $\beta$ as low as we have found for the dpAGB sources; their lowest value is $\beta=0.87$ for cold 3\,cm grains, for which
they give $\chi_{\nu}(0.85\,mm)=0.23$ cm$^{2}$g$^{-1}$. 

Significantly larger values of the dust emissivity at $\sim1$\,mm have been used in the literature, e.g., Jura et al.
(1997) assume $\chi_{\nu}(1.3mm)=3$ cm$^{2}$g$^{-1}$ for the carbon-rich grains in the PPN, the Red Rectangle.
Dasyra et al. (2005) find $\chi_{\nu}(0.85\,mm)=1.4$ cm$^{2}$g$^{-1}$ from detailed modeling of the SED of 3 spiral
galaxies. Quoting similar results from studies of (i) high-latitude Galactic interstellar regions by del Burgo et
al. (2003), (ii) protostellar-core grains with ice mantles by Ossenkopf \& Henning (1994), and (iii) the spiral galaxy
M51 by Meijerink et al. (2005), they argue for an enhanced emissivity at submm wavelengths.  We derive conservative mass
estimates, adopting $\chi_{\nu}(0.85\,mm)=1.4$\,cm$^{2}$g$^{-1}$. The dust
temperature is assumed to be $T_d=50$\,K for IRAS22036 (based on the detailed modelling by Sahai et al. 2006), and
$T_d=150$\,K for the dpAGB sources (based on the detailed modelling by Gielen et al. 2007). The derived dust masses are
inversely proportional to the dust emissivity and temperature, hence easily scaled to different values of these parameters.

In summary, our pilot EVLA survey confirms our hypothesis of large grain emission for two key pAGB objects, and thus
it is likely to be applicable to these as a class -- however, much more sensitive surveys of larger target
samples at submm/mm and cm
wavelengths are needed to establish it definitively. Current and upcoming facilities such as CARMA, ALMA and the EVLA
will allow such surveys to be carried out with fairly modest integration times. 
We have proposed new observations with
the EVLA which already allows a factor\,8 larger continuum bandwidth ($2\times1$\,GHz compared to $2\times128$\,MHz in
our pilot survey), and will be significantly larger once the EVLA project is completed.

RS's contribution to the
research described here was carried out at the Jet Propulsion Laboratory, California Institute of Technology, under a
contract with NASA. Financial support was provided by NASA through a Long Term Space Astrophysics award (to RS and MM).

\clearpage
\begin{table}
\caption{Radio and Millimeter-Wave Fluxes of post-AGB Objects}
\begin{tabular}{l c c c c c c c c}
\hline\noalign{\smallskip}
Source  & X  & Ka  & Q & 3\,mm\tablenotemark{a} & 1.3\,mm\tablenotemark{b} & 0.85\,mm & D\tablenotemark{c} & $M_d$\\ 
& $\mu$Jy($\sigma$)&$\mu$Jy($\sigma$)& $\mu$Jy($\sigma$)& mJy($\sigma$)& mJy($\sigma$) & mJy($\sigma$)& kpc & $10^{-2}$\ms\\
\hline\noalign{\smallskip}
RV\,Tau         &   ...  &  270\,(50)& (107) & 3.9(0.2) & ... &  50.3\,(3.6)\tablenotemark{d} & 2.2 & 0.1 \\ 
U\,Mon          &  ...  &  (100)    & (169) & 15(0.3)  & 100(14) & 182\,(2.6)\tablenotemark{d} & 0.77 & 0.064 \\ 
AC\,Her         &    (46)  &   ...  &  ...  & 4.6(0.4) & 38(1) & 99.4\,(3.8)\tablenotemark{d} & 1.1 & 0.072\\
IRAS16342$-$3814 &   (162) &   (168)   &  (254) & ... & 277\,(13)\tablenotemark{e}  & 602\,(90)\tablenotemark{f} & & \\ 
IRAS17150$-$3224 &    ...  &   (240)    &  (213) & ... & 158\,(10)\tablenotemark{e}  & ... & & \\
IRAS18135$-$1456 &    (66) &    (82)    &  (169) & 12\,(1.4)\tablenotemark{g} & ...  & ... & & \\
IRAS18276$-$1431 &    ...  &    (108)   &  (157) & 11\,(3.2)\tablenotemark{g} & ...  &  ... & & \\
IRAS19548+3035 &    (45)   &    ...    &  ...  & 6\,(1.1)\tablenotemark{g} & ...  &  ... & & \\
IRAS20000+3239&(44)& ...& ... & 6\,(1.1)\tablenotemark{g} & 11.4(1.7)\tablenotemark{h} & 30.9\,(2.5)\tablenotemark{i} & &\\
IRAS22036+5306 & 1010\,(62) & 1180\,(55) & 1230\,(81)& 8.4\,(0.7)\tablenotemark{j} & ...  & 290\,(40)\tablenotemark{j} & 2 & 2.2\\
\hline\noalign{\smallskip}
\end{tabular}
\label{fluxes}
\tablenotetext{a}{Beam sizes for RV\,Tau, U\,Mon, \& AC\,Her 3\,mm observations are 2\farcs4$\times$1\farcs5, 
2\farcs4$\times$2\farcs1, \& 2\farcs4$\times$1\farcs5, respectively}
\tablenotetext{b}{Beam sizes U\,Mon \& AC\,Her 1.3\,mm observations are 2\farcs2$\times$0\farcs9 \& 2\farcs0$\times$1\farcs8,
respectively}
\tablenotetext{c}{Distances for RV\,Tau, U\,Mon, \& AC\,Her from de Ruyter et al. (2005), for IRAS22036 from Sahai et al. (2003)}
\tablenotetext{d}{de Ruyter et al. 2005}
\tablenotetext{e}{G\"urtler et al. 1996}
\tablenotetext{f}{Ladjal et al. 2010, central wavelength 0.87\,mm} 
\tablenotetext{g}{S{\'a}nchez Contreras \& Sahai 2011, central wavelength 2.6\,mm} 
\tablenotetext{h}{Buemi et al. 2007, central wavelength 1.2\,mm} 
\tablenotetext{i}{Gledhill et al. 2002} 
\tablenotetext{j}{Sahai et al. 2006, central wavelengths 2.6\,mm and 0.88\,mm}                                                                                         
\end{table}


\begin{thebibliography}{}
\bibitem[Balick et al.(2002)]{bf02} Balick, B.~\& Frank, A.\ 2002, Ann.Rev.Astr.Ap., 40, 439
\bibitem[Buemi et 
al.(2007)]{2007A&A...462..637B} Buemi, C.~S., Umana, G., Trigilio, C., \& Leto, P.\ 2007, \aap, 462, 637
\bibitem[Bujarrabal et al.(2003)]{buj03} Bujarrabal, V. et al. 2003, A\&A, 409, 573
\bibitem[Cohen et al.(2004)]{2004AJ....127.2362C} Cohen, M., Van Winckel, 
H., Bond, H.~E., \& Gull, T.~R.\ 2004, \aj, 127, 2362 
\bibitem[Dasyra et 
al.(2005)]{2005A&A...437..447D} Dasyra, K.~M., Xilouris, E.~M., Misiriotis, A., \& Kylafis, N.~D.\ 2005, \aap, 437, 447
\bibitem[de Ruyter et al.(2005)]{deruyt05} de Ruyter, S., Van Winckel, H., Dominik, C., et al. 2005, A\&A, 435, 161
\bibitem[de Ruyter et al.(2006)]{deruyt06} de Ruyter, S., Van Winckel, H., Maas, T. et al. 2006, A\&A, 448, 641
\bibitem[Draine(2006)]{2006ApJ...636.1114D} Draine, B.~T.\ 2006, \apj, 636, 
1114 
\bibitem[del Burgo et al.(2003)]{2003MNRAS.346..403D} del Burgo, C., 
Laureijs, R.~J., {\'A}brah{\'a}m, P., \& Kiss, C.\ 2003, \mnras, 346, 403 
\bibitem[Gielen et 
al.(2007)]{2007A&A...475..629G} Gielen, C., van Winckel, H., Waters, L.~B.~F.~M., Min, M., \& Dominik, C.\ 2007, \aap,
475, 629
\bibitem[Gledhill et al.(2001)]{gled01} Gledhill, T.~M., Chrysostomou, A., Hough, J.~H., \& Yates, J.~A. 2001, MNRAS, 322, 321
\bibitem[G\"urtler et al.(1996)]{gurt96} G\"urtler, J., K\"ompe, C. \& Henning, Th. 1996, A\&A, 305, 878 
\bibitem[Ivezic, Nenkova, \& Elitzur(1999)]{ivezic99} Ivezic, Z., Nenkova, \& Elitzur, M. 1999,
  University of Kentucky Internal Report
\bibitem[Jura et al.(1997)]{jura97} Jura, M., Turner, J., \& Balm, S.~P.\ 1997, ApJ, 474, 741
\bibitem[Jura et al.(2000)]{jura00} Jura, M., Turner, J.~L., Van Dyk, S., \& Knapp, G.~R.\,2000, ApJ, 528, L105
\bibitem[Ladjal et al.(2010)]{ladj10} Ladjal, D., Justtanont, K., Groenewegen, M.~A.~T., Blommaert, J.~A.~D.~L., Waelkens, C.,
\& Barlow, M.~J.\ 2010, \aap, 513, A53
\bibitem[Maas et al.(2005)]{2005A&A...429..297M} Maas, T., Van Winckel, H., \& Lloyd Evans, T.\ 2005, \aap, 429, 297
\bibitem[Mastrodemos(1998)]{mast98} Mastrodemos, M. \& Morris, M. 1998, ApJ 497, 303
\bibitem[Meijerink et 
al.(2005)]{2005A&A...430..427M} Meijerink, R., Tilanus, R.~P.~J., Dullemond, C.~P., Israel, F.~P., \& van der Werf,
P.~P.\ 2005, \aap, 430, 427 
\bibitem[Moe(1999)]{moe99} Moe,  M. \&   De Marco, O.,  2006. ApJ, 650, 916
\bibitem[Morris(1987)]{mor87} Morris, M.\ 1987, PASP, 99, 1115
\bibitem[Ossenkopf 
\& Henning(1994)]{1994A&A...291..943O} Ossenkopf, V., \& Henning, T.\ 1994, \aap, 291, 943 
\bibitem[Ossenkopf et 
al.(1992)]{1992A&A...261..567O} Ossenkopf, V., Henning, T., \& Mathis, J.~S.\ 1992, \aap, 261, 567
\bibitem[Pollack et al.(1994)]{poll94} Pollack, J.~B., Hollenbach, D., Beckwith, S., Simonelli, D.~P., Roush, T., \& Fong, W.\
1994, ApJ, 421, 615
\bibitem[Reyes-Ruiz(1999)]{rey99} Reyes-Ruiz, M., \& Lopez, J. A. 1999 ApJ, 524, 952
\bibitem[Sahai(1998)]{sah98} Sahai, R. \& Trauger, J.T. 1998, AJ, 116, 1357
\bibitem[Sahai et al.(2003)]{2003ApJ...586L..81S} Sahai, R., Zijlstra, A., 
S{\'a}nchez Contreras, C., \& Morris, M.\ 2003, \apjl, 586, L81 
\bibitem[Sahai et al.(2006)]{sah06} Sahai, R., Young, K., Patel, N., S{\' a}nchez Contreras, C., \& Morris, M.\ 2006, ApJ, 653, 1241
\bibitem[Sahai et al.(2007a)]{sah07a} Sahai, R., Morris, M., S{\'a}nchez Contreras, C., \& Claussen, M.\ 2007a, AJ, 134, 2200
\bibitem[Sahai et al.(2007b)]{sah07b} Sahai, R., S{\'a}nchez Contreras, C., Morris, M., \& Claussen, M.\ 2007b, ApJ, 658, 410
\bibitem[Sahai et al.(2011)]{2011AJ....141..134S} Sahai, R., Morris, M.~R., 
\& Villar, G.~G.\ 2011, \aj, 141, 134 
\bibitem[Contreras et al.(2007)]{sc07} S{\'a}nchez Contreras, C., Le Mignant, D., Sahai, R., Gil de Paz, A., 
\& Morris, M.\ 2007, ApJ, 656, 1150 
\bibitem[Contreras et al.(2007)]{sc08} S{\'a}nchez Contreras, C., Sahai, R., Gil de Paz, A., \& Goodrich, R.\ 2008, ApJS, 179, 166
\bibitem[Contreras et al.(2011)]{sc11} S{\'a}nchez Contreras, C. \& Sahai, R. 2011, ApJS (submitted)
\bibitem[Soker(2004)]{} Soker, N., \& Livio, M., 2004,  ApJ, 421, 219
\bibitem[Van Winckel et al.(2008)]{vw08} Van Winckel et al. \ 2008, AIP Conf. Proc., 1001, 349
\bibitem[Waters et al.(2008)]{watr08} Waters, L.B.F.M., Trams, N.R., \& Waelkens, C.\ 1992, A\&A, 262, L37
\bibitem[Wright \& Barlow(1975)]{1975MNRAS.170...41W} Wright, A.~E., \& Barlow, M.~J.\ 1975, \mnras, 170, 41                     
\end{thebibliography}
\end{document}